# Evaluation of Variability Concepts for Simulink in the Automotive Domain

Carsten Kolassa
Software Engineering
RWTH Aachen University
kolassa@se-rwth.de

Holger Rendel
Volkswagen AG
holger.rendel@volkswagen.de

Bernhard Rumpe
Software Engineering
RWTH Aachen University
http://www.se-rwth.de

**Abstract**

*Modeling variability in Matlab/Simulink becomes more and more important. We took the two variability modeling concepts already included in Matlab/Simulink and our own one and evaluated them to find out which one is suited best for modeling variability in the automotive domain. We conducted a controlled experiment with developers at Volkswagen AG to decide which concept is preferred by developers and if their preference aligns with measurable performance factors. We found out that all existing concepts are viable approaches and that the delta approach is both the preferred concept as well as the objectively most efficient one, which makes Delta-Simulink a good solution to model variability in the automotive domain.*

## 1. Introduction

The configurations of modern cars are highly variable. This variability concerns all aspects of the vehicle functionality from technical base functionality to comfort functionality like vehicle speed based steering support. Managing this variability is essential. Customers wish to tailor their car to their own preferences, e.g. in regard to comfort, safety, or sportiness. This often leads to sport and limousine versions of one and the same car. Another factor is the reuse of identical hardware parts within one corporate group with different software. Audi and Porsche are, for example, part of the Volkswagen group. The Audi A6 and the Porsche Macan, therefore, use a similar steering box assembly, the difference is mainly in the software they use. However, in order to realize this variability, the different variants of the vehicle functions have to be planned and realized during development. This requires to be able to handle variability in all development phases by adequate means. To manage this variability and the potential reuse of components, Software Product Line Engineering (SPLE) methods are used. In the automotive domain, functions of a system are often developed in Matlab/Simulink [3], which offers built in support to manage variants. Since 2007 Simulink itself has been providing an annotative variability modeling approach [17] where a model contains all Simulink blocks that may be contained in any variant such that it is also called a 150%-Model. Simple variant management can be done by using conditional blocks like if, switch, and action blocks that are either active or inactive. Another annotative approach using variant subsystems was introduced to Simulink in release R2009b and R2010b [19]. These two concepts are limited and a new variability modeling method for Simulink was developed which is based on delta modeling [13]. This new concept of variability in Simulink, is called Delta-Simulink [5]. We want to compare 150%-models with a transformational variability approach in Simulink to answer the question which modeling approach is suited best with the focus being on the automotive domain. In order to answer this question, we conduct a controlled experiment as well as a survey with developers at Volkswagen. The contribution of this paper is an in-depth comparison of variability modeling concepts in Simulink. The concepts are compared by a set of tasks and scenarios. In an evaluation at an OEM (original equipment manufacturer) all concepts are measured and validated in a real development situation.

The paper is structured as follows: In Sect. 2, we describe the concepts used to model variability in Simulink. We explain in Sect. 3 how we designed the study to compare the variability concepts. In Sect. 4 we show the results of our study. Threats to validity of our study are discussed in Sect. 5. Sect. 6 reviews related work and Sect. 7 concludes this paper.

## 2. Variability Concepts

In the following section, we give an overview of the variability concepts we evaluate.

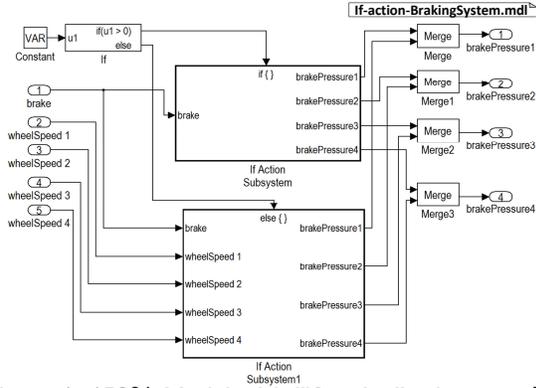

Figure 1: 150%-Model with "If-action" subsystem [5]

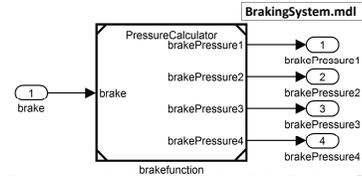

Figure 2: Base model as used with Delta-Simulink [5]

## 2.1. 150%-Models

The already available concepts to model variability in Simulink use an annotative variability modeling approach [17] where all possible variants are described in the same model. The models contain all Simulink blocks that may be used in a variant hence these concept are also called 150%-Models. In the following we describe the two available concepts to create 150%-Models. This differentiation was also used by the evaluation in [19].

**150%-Models with Conditional Model Elements (CME).** In models with conditional model elements each variant is encapsulated in a separate subsystem as shown in Figure 1. Each of these "If-action"-subsystems contains an extra activation port to enable or disable it. In addition an extra "if-block" or "switch-block" evaluates a condition based on a configuration variable. The outputs are connected to the activation port of an "If-action"-subsystem. As the subsystem blocks generate the same signal a merge block is necessary for each output signal.

**150%-Models with Model Elements for Model Adaptation (MEMA).** This approach uses special blocks which are variation points for subsystems or model references. They contain inputs for all possible input signals over all variants and outputs for all possible output signals. Inside the variant subsystem, each variant is encapsulated in a single subsystem. The active variant is determined by conditions that are attached to each block and the corresponding connections are created during the simulation. Only one variant can be active.

## 2.2. Transformational Models

Transformational variant modeling approaches modify a base system to build variants. In the following, the own-developed Simulink extension Delta-Simulink is described.

**Delta-Simulink.** Delta-Simulink introduced in [5] is a transformational approach for modeling software variability. This variability modeling mechanism consists of deltas which are applied to a base model. The deltas are Simulink models where elements may be annotated with delta operations. These operations are encoded by colours in a separate delta model view. Therefore, it is possible to use the colours as before in the "normal view". Green elements are added in the delta, red elements are removed, orange elements are replaced and blue elements are modified (for example, subsystems). A specific variant can be obtained by applying a series of deltas to the base model. The order of the deltas can be arranged with application order constraints. To make this more clear, we go through an example step by step. In our example, we have the base functionality given in Figure 2. The variant is modeled in a separate model file, the delta model. In our case, we want to replace the model reference of the brakefunction with the reference for the ABS function and we want to add the needed ports. In order to model this variant, we add the reference to the ABS model to our delta model and annotate it with a replace operation. We add the ports and connections and annotate them with the add operation. The delta, for our example, is shown in Figure 3.

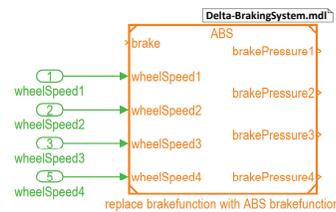

Figure 3: Delta-Simulink model (The add operations are marked green and the replace operation is marked orange.) [5]

The delta can then be applied to the base model and the variant with ABS functionality is generated. A more detailed description of Delta-Simulink and its functionality is given in [5].

For this evaluation, we used a version of Delta-Simulink that is able to model deltas for arbitrary blocks and not just for models or subsystems as in the original publication.

## 3. Study Design

This section explains how we designed the study and the goals we want to achieve. The study consists of a survey and a controlled experiment.

We designed the study to answer the following research questions:

**RQ1.1: With which concept can developers model variability the fastest?** A concept is chosen if it takes a minimum of time to apply it.

**RQ1.2: Which concept is accepted best by the developers?** A concept is accepted if the developers prefer one concept to the other ones and actually want to use it in their daily work.

The answer is also derived from sub questions.

**RQ2.1: Which approach yields the smallest models?** Small models are better to maintain. Variability concepts with a lower quantity of blocks and connections are less error-prone upon modification.

**RQ2.2: In which concept are the features encapsulated best?** Encapsulation means separating the variable parts (e.g. in own artifacts, or in own blocks) to clarify the variability and to support reuse of features in other contexts.

**RQ2.3: Which concept yields the most well structured models?** Well structured models are easy to read. This is what we want to assess with "clearness". This supports the developer to comprehend the model and to identify the variability.

### 3.1. Self-assessment and Introduction to the Concepts

At the beginning of the study the developers self assess their experience in model based development. Then they watch a video that explains the three modeling concepts. After that the developers are asked to self assess their experience with Simulink and to decide how easily understandable the three concepts are. The study participants have been instructed to rate the understandability of the concepts here not how well the concepts are presented in the video.

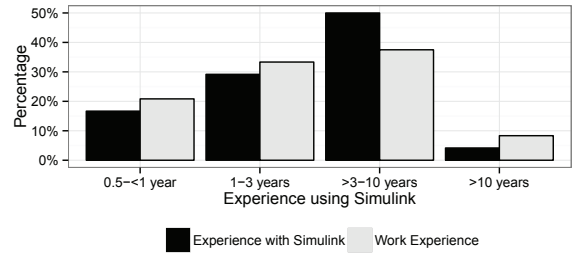

Figure 4: The experience of the participants

We tried to get a good representation of the developer population at Volkswagen and took a random sample of two divisions to achieve this. It wasn't possible to check whether this random sample is a perfect representation of the overall developer population as the privacy policy of Volkswagen made it impossible to measure it. Figure 4 shows the work experience as well as the Simulink experience of our study participants.

### 3.2. Objective Assessment

The second part of the study consists of an objective assessment as a controlled experiment. In the controlled experiment the participants had to complete four tasks which were given in the form of change requests (CRs). This is the standard format of tasks at the surveyed divisions of Volkswagen AG.

**3.2.1. How we Chose the Tasks for the Experiment.** In order to make our results transferable to other settings in the automotive domain as well as to make them meaningful within the tested setting and to prevent the introduction of biases we attached great importance to choosing the tasks for the controlled experiment.

We didn't choose the tasks completely ourselves as this could introduce a selection bias (e.g. we could choose tasks that benefit one particular modeling concept). Instead we talked with three domain experts at Volkswagen that did not know the modeling concepts and asked them about potential tasks. Together with these domain experts we analysed past tasks in the change management system. We looked for past tasks that would also make sense as variable features in a product line.

Two tasks have been directly derived from previous tasks, while the third represents a future task in an actual project and the fourth one was constructed.

The Tasks/CRs are the following:

**Task 1:** Adding a filter for an existing output port. This requires the addition of a filter block which is placed before the output port. Two variants, one with filter and one without filter, should be supported.

**Task 2:** Changing several fixed factors to different lookup tables in subsystems. For this, an additional input must be added to parameterize the lookup tables. Two different subsystems must process the new input and be modified by adding lookup tables which use the new input. The resulting variants are one with lookup tables and one without lookup tables.

**Task 3:** Changing the output data type of a port while adding a logging mechanism in another part of the model. This task requires similar actions as Task 1 as well as an output port addition. This task modifies different model parts. Thus, the corresponding variants must cover both changes simultaneously or none of them.

**Task 4:** Reading an existing signal from a bus and adding new functionality for this signal including a new output. This change needs the modification of Simulink properties of an existing model part as well as the addition of the new port. The variants are one with the realization of the task and one without its realization. In case the tested concept does not allow interface changes the study participants were allowed to put a ground signal on a port in the variant the port is not used.

In order to further validate our choice of tasks we asked the participants within the study itself whether the task is representative for the automotive domain. We present these results in section 5.

**3.2.2. Data Analysis.** Our experimental setup is a *Cross balanced repeated measures within subject design* [16] which means that we use the same subjects in every branch of our experiment. Each participant repeats the same tasks with every method. In order to prevent biases, like a learning effect in which a study participant is faster when he does the same task a second time, we use counterbalanced measures. In this kind of experiment design the order in which the concepts have to be executed changes between subjects. We evaluate 3 concepts which leads to 6 different possible orders and therefore 6 groups each with a different order. Each group consists of 4 developers giving us a sample size $N = 24$.

For each task we measure the following variables:

1) Time and effort (mouse movement) the task takes.
2) Number of blocks in the resulting model.

Modeling a variation consists of two different kinds of work, the manual work (mouse movements and clicks) and the mental work (creating the model in one's mind). Time accounts for manual and mental work while mouse movements only accounts for manual work.

### 3.3. Subjective Assessment

Our study participants have a considerable experience in the automotive domain and in using Simulink. In order to choose the most appropriate method for modeling variability, we asked the participants for a subjective assessment of the concepts.

The study participants have been asked to answer the following questions for each concept:

**SQ1:** How easy to understand are the concepts? This describes how easy it is for a developer to get familiar with the concept in case he has never used it.
**SQ2:** How clear are the models of the concepts? The participants have been instructed that they should evaluate how easy it is to spot important parts of the model.
**SQ3:** How well do the concepts encapsulate the features? The participants have been instructed that they should evaluate how well the variable parts are separated from the rest of the model.
**SQ4:** How well is the modeling concept suited for the automotive domain?

The exact definition of what we mean with encapsulation or clearness as explained in our research questions has been made clear to the participants by using examples. They have been encouraged to ask questions in case one of the terms was unclear.

We are aware that this subjective assessment on its own is not an adequate measurement to make the final decision on which concept is suited best. It is merely a second validation of our objective measurements in the controlled experiment.

### 4. Results

In the following section we present the results of our study.

### 4.1. Results of the Objective Assessment

We start with the objective assessment of the modeling concepts.

**4.1.1. Time Measurements.** In order to answer **RQ1.1** we need to test the following hypothesis for every concept:

**H1.1:** In concept X the tasks can be fulfilled faster.

Where X is one of the concepts we evaluate.

In the objective assessment we measure the time each participant needs for each of the tasks. We use each task-subject pair as a data point.

We analyse our data using a linear mixed model as described in [7], [18]. We decided against using a repeated measures analysis of variance (rAnova) on the data because rAnova is vulnerable against unequivalent time points and we would need to assume sphericity or compound symmetry [11].

The linear mixed model does not have these problems but like rAnova it adjusts for the between subjects error.

This between subjects error is for example the difference between two developers A and B which is independent of the modeling concepts. If subject A is always faster by a constant factor than subject B then this difference is not caused by the modeling concepts but by the subjects themselves. Linear mixed models account for this error and it is excluded to yield results independent of the differences between individual subjects.

We fit a linear mixed model for each task over all concepts. We got significant results for all tasks except for the influence of the MEMA concept on Task 2 (see first block of Table 1). Significant results in our case mean that there is a statistically measurable difference with a confidence level of at least 95%.

We are interested in how big this influence is and therefore conducted a post-hoc analysis of our results using the Tukey HSD Test [6]. The time row of Table 2 shows the time differences needed to complete each task adjusted for between subject errors.

We test for the Null-Hypothesis given in Table 3 that the difference between each concept pair is smaller than 0 and calculated the p-values for each task. They are close to zero which means that the difference is bigger than zero with statistical significance. With one exception for Task 2 we see only the qualitative result that the hypothesis can be rejected but it cannot be rejected with statistical significance.

**Table 1: P-values for the insignificance of the factors in the linear mixed model**

|        |       | Task1  | Task2  | Task3  | Task4  |
|--------|-------|--------|--------|--------|--------|
| Time[s]| Delta | 0.0008 | 0.0000 | 0.0000 | 0.0000 |
|        | MEMA  | 0.0132 | 0.1172 | 0.0002 | 0.0007 |
|        | CME   | 0.0000 | 0.0001 | 0.0000 | 0.0000 |
| Pixel  | Delta | 0.0002 | 0.0000 | 0.0000 | 0.0000 |
|        | MEMA  | 0.0173 | 0.0073 | 0.0004 | 0.0001 |
|        | CME   | 0.0000 | 0.0000 | 0.0000 | 0.0001 |
| Blocks | Delta | 0.0000 | 0.0000 | 0.0000 | 0.0000 |
|        | MEMA  | 0.0083 | 0.0071 | 0.0000 | 0.0000 |
|        | CME   | 0.0153 | 0.2217 | 0.0000 | 0.0000 |

**Table 3: P-values for linear hypothesis**

|        | Hypothesis | Task1 | Task2 | Task3 | Task4 |
|--------|-----------|-------|-------|-------|-------|
| Time[s]| MEMA - Delta $\leq 0$ | 0.0138 | 0.1371 | 0.0001 | 0.0004 |
|        | CME - Delta $\leq 0$ | 0.0000 | 0.0000 | 0.0000 | 0.0000 |
|        | CME - MEMA $\leq 0$ | 0.0000 | 0.0137 | 0.0512 | 0.2569 |
| Pixel  | MEMA - delta $\leq 0$ | 0.0184 | 0.0068 | 0.0001 | 0.0000 |
|        | CME - Delta $\leq 0$ | 0.0000 | 0.0000 | 0.0000 | 0.0000 |
|        | CME - MEMA $\leq 0$ | 0.0012 | 0.0702 | 0.2836 | 0.8260 |
| Blocks | MEMA - delta $\leq 0$ | 0.0080 | 0.0069 | 0.0000 | 0.0000 |
|        | CME - Delta $\leq 0$ | 0.0165 | 0.2509 | 0.0000 | 0.0000 |
|        | CME - MEMA $\leq 0$ | 0.9129 | 0.9999 | 0.5424 | 0.0617 |

We are also interested in the absolute numbers of this difference and calculated the estimates for this difference and the corresponding error margins (see Table 2).

This means that our initial hypothesis **H1.1** is true for the Delta-Simulink approach in 3 out of 4 tasks to the 95% confidence interval while we see positive but not statistically significant results in Task 2.

**4.1.2. Measuring Mouse Movements.** We also measured the mouse movements of the study par-ticipants in pixel. The mouse movements in pixel is a performance measurement like the time measurements but it is independent of thinking processes (the time the participant needs to think to come up with a solution) and only accounts for the manual steps needed to model the solution. We measured it as it shows, whether most time is spent doing manual work (moving the mouse) or thinking.

**Table 2: Linear mixed model estimates with confidence intervals to the 95% confidence level**

|        |              | Task1 | | | Task2 | | | Task3 | | | Task4 | | |
|--------|--------------|----------|----------|----------|----------|----------|----------|----------|----------|----------|----------|----------|----------|
|        |              | Estimate | lwr | upr | Estimate | lwr | upr | Estimate | lwr | upr | Estimate | lwr | upr |
| Time[s]| MEMA - Delta | 67.83 | 6.16 | 129.51 | 47.71 | -22.32 | 117.74 | 84.61 | 36.34 | 132.88 | 60.65 | 21.57 | 99.74 |
|        | CME - Delta  | 209.71 | 148.03 | 271.38 | 124.83 | 54.80 | 194.87 | 127.88 | 78.97 | 176.79 | 81.04 | 41.96 | 120.13 |
|        | CME - MEMA   | 141.88 | 80.20 | 203.55 | 77.13 | 7.09 | 147.16 | 43.27 | -5.64 | 92.18 | 20.39 | -18.69 | 59.48 |
| Pixel  | MEMA - delta | 27906.61 | 1495.69 | 54317.53 | 31350.41 | 5298.22 | 57402.60 | 32780.64 | 13007.37 | 52553.91 | 28936.86 | 13769.89 | 44103.84 |
|        | CME - Delta  | 66040.73 | 39320.95 | 92760.51 | 52811.55 | 26759.35 | 78863.74 | 42535.27 | 22762.00 | 62308.54 | 29054.95 | 13887.98 | 44221.93 |
|        | CME - MEMA   | 38134.12 | 11414.34 | 64853.90 | 21461.14 | -4591.06 | 47513.33 | 9754.64 | -10018.63 | 29527.91 | 118.09 | -15048.88 | 15285.06 |
| Blocks | MEMA - delta | 84.54 | 12.75 | 156.33 | 86.50 | 14.52 | 158.48 | 14.61 | 12.61 | 16.61 | 10.09 | 8.75 | 11.42 |
|        | CME - Delta  | 77.17 | 5.38 | 148.96 | 38.04 | -33.94 | 110.02 | 15.13 | 13.13 | 17.13 | 11.22 | 9.89 | 12.55 |
|        | CME - MEMA   | -7.38 | -79.16 | 64.41 | -48.46 | -120.44 | 23.52 | 0.52 | -1.48 | 2.53 | 1.13 | -0.20 | 2.46 |

We used the same statistical methods to analyse the results as in subsubsection 4.1.1. The p-values for the significance of each factor are shown in the second block of Table 1. We also conducted a post hoc analysis, we tested against the null hypothesis that the difference between the concepts is lower than zero as shown in Table 3. The p-values for this hypothesis are close to zero, which means that we can reject it and that the difference is in fact positive with statistical significance.

The mouse movements show statistically clearer results as the time measurements. Here, we can conclude that delta modeling needs less mouse movements for every single task. That means, in comparison to the time measurements, that more time was spent doing manual work in the CME and MEMA approaches. The estimates for the difference are given in the pixel row of Table 2.

### 4.1.3. Measuring the Size of the Resulting Models.
We used the same approach for the analysis of the model size as for the time and the mouse movements. The analysis answers **RQ2.1** and we test for the hypothesis:

**H2.1:** In concept X the resulting models have the smallest number of blocks.

The p-values for the significance of each factor are shown in the third block of Table 1. Our post hoc analysis showed that Delta-Simulink has the smallest model size estimates for all 4 tasks (see Table 2) but statistical significance could only be achieved in 3 of them (see Table 1). The model size is an indicator for the clearness of a model. For Delta-Simulink, the number of blocks includes the blocks of the base model as well as the blocks of the delta model, that means the number of blocks of the delta model itself is even smaller as it only contains the difference to the base model.

## 4.2. Results of the Subjective Assessment

In our study we had the privilege to work with subjects that have experience in the automotive domain. This means that the subjects themselves can look at the modeling concepts and give us their educated view on how they perform.

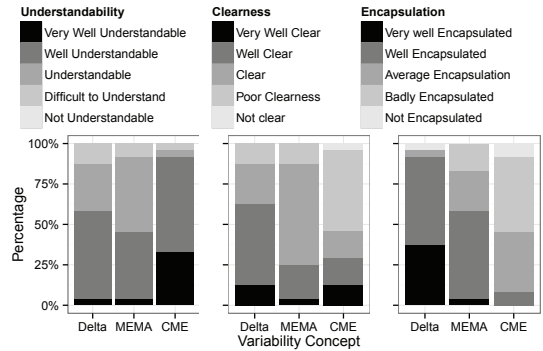

**Figure 5: Subjective assessment along various variables**

**4.2.1. Assessment of the Concepts.** We presented several models to our subjects and had them evaluate the concepts according to understandability, clearness, and encapsulation as described in subsection 3.3. This allows to test the following hypotheses which answer **RQ2.2** and **RQ2.3**:

**H2.2** For one concept X the feature encapsulation is regarded better by the subjects than for the others.

**H2.3** For one concept X the models are regarded as clearer than for the others.

The results of this evaluation are shown in Figure 5 and correspond to **SQ1-SQ3**. The Delta-Simulink approach leads in clearness and encapsulation while the CME approach is the easiest to understand.

In order to answer **H2.2** and **H2.3** with statistical significance we used the Wilcoxon-Mann-Whitney two-sample rank-sum test [9] and compared the concepts pairwise. We chose this test as our data is pairwise (the data pairs come from the same subject who grades the concepts on the same scale) and ordinal [10].

We tested against the following hypotheses:

$H_{0a} : Y_1 - Y_2 = 0;$   $H_{0b} : Y_1 - Y_2 > 0;$
$H_{1a} : Y_1 - Y_2 \neq 0;$   $H_{1b} : Y_1 - Y_2 \leq 0;$

In the first test the null hypothesis is $H_{0a}$, i.e. the median of the assessment is equal to the one of the other concept. The second Null-Hypothesis is $H_{0b}$, i.e. the difference between the medians of the assessments of two concepts is bigger than 0. In this case, the second concept would perform worse than the first, as a higher value means a better rating.

Table 4 shows the p-values of our hypotheses. All the p-values for the pairs with Delta-Simulink are close to zero and smaller than the significance level of 0.05 which means that we can reject $H_{0a}$ and $H_{0b}$, thus the alternatives $H_{1a}$ and $H_{1b}$ are true we can therefore validate **H2.2** and **H2.3** for Delta-Simulink.

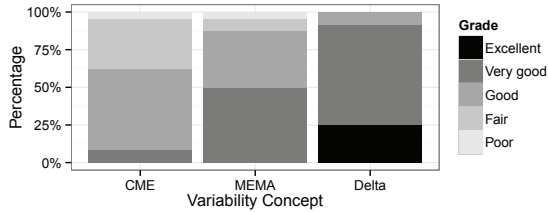

**Figure 6: Comparison of the suitability grades for the tested variability Concepts**

**Table 4: Results Wilcoxon-Mann-Whitney Two-Sample Rank-Sum for Clearness and Encapsulation**

|  | Hypothesis | $H_{0a}$ | $H_{0b}$ |
|---|---|---|---|
|  | CME - Delta | 2.361E-02 | 1.181E-02 |
| Clearness | MEMA - Delta | 4.664E-02 | 2.332E-02 |
|  | CME - MEMA | 2.644E-01 | 1.322E-01 |
|  | CME - Delta | 4.692E-05 | 2.346E-05 |
| Encapsulation | MEMA - Delta | 1.250E-03 | 6.249E-04 |
|  | CME - MEMA | 3.378E-04 | 1.689E-04 |

**4.2.2. Overall Impression of the Subjects.**
The developers are the people who have to work with the modeling concept. Their impression is, therefore, important as a successful concept needs to be accepted by the developers. In order to answer **RQ1.2**, we need to test the following hypothesis:

**H1.2:** Concept X is better accepted by the developers than the other concepts.

We asked our subjects to rate the modeling concepts with regard to suitability in the automotive domain (**SQ4**) with grades from excellent to very poor (1 = Excellent; 2 = Very good; 3 = Good; 4 = Fair; 5 = Poor; 6=Very poor). The results are presented in Figure 6.

We used the Wilcoxon-Mann-Whitney two-sample rank-sum test again with a similar null hypotheses as in the previous section. With one difference the second hypothesis is not $H_{0b}$ but $H_{0c}$ as in this case a lower rating is the better one.

$H_{0c} : Y_1 - Y_2 < 0;$     $H_{1c} : Y_1 - Y_2 \geq 0;$

Table 5 shows that the p-values are again small which means that we can reject both null hypothesis. Thus $H_{1a}$ and $H_{1c}$ are true, which means that the median grade for Delta-Simulink is smaller and therefore better than for the other concepts thereby validating **H1.2**.

**Table 5: Results Wilcoxon-Mann-Whitney Two-Sample Rank-Sum**

|  | Hypothesis | $H_{0a}$ | $H_{0c}$ |
|---|---|---|---|
| 1 | CME - Delta | 2.285E-03 | 1.143E-03 |
| 2 | MEMA - Delta | 2.209E-04 | 1.105E-04 |
| 3 | CME - MEMA | 2.285E-03 | 1.143E-03 |

### 4.3. Summary

In general the Delta approach is faster, needs less mouse movements and leads to smaller models and was better accepted by the subjects.

The only metric which was rated less positive was understandability. In our opinion that is not that problematic, as it only suggests that delta modeling has a steeper learning curve then the two other concepts. In the following, we answer the research questions to give a short summary.

**RQ1.1: Which concept enables the developers to model variability the fastest?**
**Answer:** The tasks could be fullfilled the fastest using Delta-Simulink. The results are statistically significant for Task 1, Task 3, and Task 4. For Task 2 we still get the result that it is likely that Delta-Simulink is the fastest yet we could not achieve statistical significance with our sample size. The differences are small but they are consistent which means that they will have a measurable cost impact as single tasks need more time. This impact is small for a single task but projects can consist of a huge number of variable features that also change over time beause of this a small but consistent difference per task will have an increasing impact over the liefetime of a project.

**RQ1.2: Which concept is accepted best by the developers?**
**Answer:** The Delta-Simulink approach is accepted best as it gets the best grading by the developers. The results are statistically significant.

**RQ1: Which of the concepts is suited best to model variability in the automotive domain, from a developer/modeling perspective?**
**Answer:** Based on the answers of **RQ1.1** we conclude that Delta-Simulink is the most efficient tool it as the developers have been able to solve the tasks the fastest and with the least mouse movements. It was also accepted best by the developers (see **RQ1.2**) and the developers found the Delta-Simulink models were the clearest.

**RQ2.1: Which approach yields the smallest models?**
**Answer:** The average of the block size of the Delta-Simulink models is the smallest. The results are statistically significant for Task 1, Task 3 and Task 4, while for Task 2 we get the qualitative result that the Delta-Simulink models are estimated to be smaller yet it cannot be shown with statistical significance as our sample was too small.

**RQ2.2: In which concept are the features encapsulated best?**
**Answer:** The developers think that the features are encapsulated the best in Delta-Simulink (statistically significant).

**RQ2.3: Which concept yields the most well structured models?**
**Answer:** The study participants think that Delta-Simulink yields the clearest models. (statistically significant)

**RQ2: Which concept has the best maintainability?**
**Answer:** Based on the answers of **RQ2.1**, **RQ2.2** and **RQ2.3**, we conclude that it is very likely that Delta-Simulink leads to models with the best maintainability as clear and small models that encapsulate variability and therefore allow reuse are key factors of maintainability.

## 5. Threats to Validity

The literature mentions four main threats to empirical research in software engineering [20]. These are threats to: Conclusion validity, Internal validity, Construct validity, External validity.

In the following section we address each threat individually.

### 5.1. Threats to Conclusion Validity

*Conclusion validity* concerns the question: "Does the treatment/change we introduced have a statistically significant effect on the outcome we measure?" [2].

We calculated the corresponding p-value for all of our statistical conclusions. We also gave error margins to a 95% confidence level, if possible. The small sample size prevented significant results in some cases. In these cases we could only show qualitative tendencies. These cases would have to be repeated with a bigger sample size to get significant results.

### 5.2. Threats to Internal Validity

*Internal validity* refers to the question: "Did the treatment/change we introduced cause the effects on the outcome? Can other factors also have had an effect?" [2]. We tried to exclude as many external influences as possible. There are two potential biases, the selection bias and the so called learning bias. The learning bias is the potential learning curve of the study participants. This is a major threat as there is the potential that the first concept tested teaches the participants something that they could use in the second or third concept. We tried to tackle this threat by the choice of our study design we had 6 groups which allowed us to change the order of the concepts in each group. This minimizes the threat that a learning bias introduces a significant external influence.

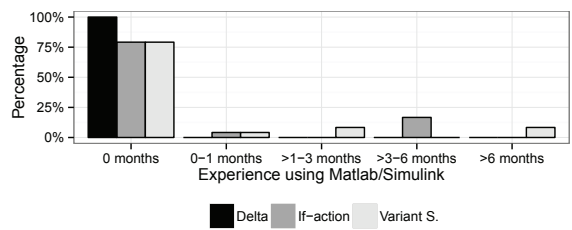

**Figure 7: The experience of the participants with the variability concepts in months**

Another threat is the previous knowledge of developers that already worked with some sort of variability concept in Simulink. Those developers likely have a bias towards that concept and they would also be better in an objective assessment as they have more experience and training in that concept. We therefore asked our study participants how much experience they had with each of the presented concepts. Figure 7 shows a summary of the answers. We don't think that the prior experience had a significant influence as the concept which leads in our objective as well as in the subjective results is the one in which all study participants had no prior experience and because a large number of participants had no experience in any of the concepts.

### 5.3. Threats to Construct Validity

*Construct validity* answers the question: "Does the treatment correspond to the actual cause we are interested in?" [2].

The study evaluates three modeling concepts for variability in Simulink. We compare them by measuring performance metrics (e.g. time used for the task, mouse movements) as well as metrics for the model size (number of blocks and connections). We also compare the subjective judgments of the participants. Our study design ensures that each participants does the same tasks and answers the same questions which leaves only the modeling concepts as a cause for the different measurements. Therefore, we believe we've achieved a considerably high construct validity for a study of this kind.

**Table 6: Representativity of the chosen tasks for the automotive domain with error margins to the 95% confidence interval**

|       |            | Not Representative | Hardly Representative | Representative | Very Representative |
|-------|------------|------|------|------|------|
| Task1 | resp. in % | 0.0  | 4.2  | 41.7 | 54.2 |
|       | error      | 0.0  | 8.2  | 20.1 | 20.4 |
| Task2 | resp. in % | 0.0  | 0.0  | 29.2 | 70.8 |
|       | error      | 0.0  | 0.0  | 18.6 | 18.6 |
| Task3 | resp. in % | 4.2  | 8.3  | 45.8 | 41.7 |
|       | error      | 8.2  | 11.3 | 20.4 | 20.1 |
| Task4 | resp. in % | 4.2  | 4.2  | 41.7 | 50.0 |
|       | error      | 8.2  | 8.2  | 20.1 | 20.4 |

### 5.4. Threats to External Validity

*External validity* concerns the question: "Is the cause and effect relationship we have shown valid in other situations?"[2]. Therefore external validity or transferability decides whether our results can be transferred to other languages or to a different setting [15].

In our case this means do our results also hold true for other Simulink models and other tasks. We tried to ensure this by talking with domain experts to get a good selection of tasks and models, we worked with external domain experts to ensure that we don't introduce a bias and we had the survey participants validate our choice by asking them how representative our choice is. Table 6 shows that in all cases more than 90% of the participants shared our notion that the chosen tasks and models are representative for the automotive domain. We also calculated errors to the 95% confidence interval which accounts for the random sampling error in our survey. Even when we assume the maximum error to the 95% confidence interval at least two thirds of our participants would still rank our tasks as representative.

### 6. Related Work

An overview of variability modeling in the solution space is given in [14]. The commonly used approaches are annotative, compositional and transformational variability modeling. Models with annotative approaches contain all variability in one model. Compositional models combine several model parts to one model variant [17]. In transformational approaches model variants are developed by transforming a base model [12]. The application of variability modeling in Simulink is discussed in [19]. Mainly annotative approaches are presented and their quality aspects, binding times, supported feature types and granularity are summarized. In contrast to our evaluation the concepts are not evaluated in a practical study. A detailed description of the Delta-Simulink approach is described in [5]. [1] shows how the concepts with standard Simulink blocks can be used with the variant management system pure:variants. This is done with variation points for binding the variable features. The application of external tools was not covered by our evaluation due to the variety of tools and concepts to bind variability outside of Matlab. [21] presents an decision-oriented approach for modeling variability in Simulink. Similar model parts are extracted and the variability is bound by decisions. The Simulink internal implementation isn't different to the CME-concept. [8] presents a concept for modeling variability in Simulink based on a 3 layered template approach. The concept enhances the usability of existing Simulink variability mechanisms by adding another layer of abstraction. It also adds an additional binding time. [4] presents another concept for modeling variability in Simulink but only a description of the concept itself is published. It aims at removing the shortcomings of variant subsystems, like problems with interface changes. It is very similar to the MEMA concept but adds a special input and output signal processing to cope with interface changes. We decided to not include [8] and [4] in our study as both are similar to the MEMA concept and because there was no implementation of the concepts available to us.

### 7. Conclusion

In this paper we compare and evaluate variability modeling concepts for Simulink in the automotive domain. Delta modeling shows very promising results in this evaluation as it leads in the subjective as well as objective measurements.

Delta-Simulink strong feature encapsulation leads to smaller models and this smaller models can be created faster and are more clear than complex 150%-Models. This evaluation is also a further validation for the approach of graphical delta modeling which was introduced with Delta-Simulink.

This study only evaluates the modeling concept which is the way the variability is presented and modeled. For the automotive domain other features are important as well like the binding time of the variability but these have not been part of this study as we solely focus on the modeling perspective.

In the future we want to conduct a second study which will compare the properties of 150%-Models and delta models further. Then we want to give a more fine grained overview on how the different concepts perform in real world scenarios.